\documentclass[aps,prb,superscriptaddress,noshowpacs, 
twocolumn
]{revtex4}
\usepackage{graphicx}
\usepackage{epstopdf}
\usepackage{amssymb,amsmath,amsfonts}

\newcommand{\coso} {{\mbox{Cu${}_2$OSeO${}_3$}}}
\begin{document}
\title{Dramatic pressure-driven enhancement of bulk skyrmion stability}
\author{I. Levati\'c}
\affiliation{Institute of Physics, Bijeni\v cka 46, HR-10 000, Zagreb, Croatia}
\author{P. Pop\v cevi\'c}
\affiliation{Institute of Physics, Bijeni\v cka 46, HR-10 000, Zagreb, Croatia}
\affiliation{Institute of Solid State Physics, TU Wien, A-1040 Vienna, Austria}
\author{V. \v Surija}
\affiliation{Institute of Physics, Bijeni\v cka 46, HR-10 000, Zagreb, Croatia}
\author{A. Kruchkov}
\affiliation{Laboratory for Quantum Magnetism, Institute of Condensed Matter Physics, EPFL, CH-1015 Lausanne, Switzerland}
\author{H. Berger}
\affiliation{Institute of Condensed Matter Physics, EPFL, CH-1015 Lausanne, Switzerland}
\author{A. Magrez}
\affiliation{Institute of Condensed Matter Physics, EPFL, CH-1015 Lausanne, Switzerland}
\author{J. S. White}
\affiliation{Laboratory for Neutron Scattering and Imaging, Paul Scherrer Institut, CH-5232 Villigen,
	Switzerland}
\author{H. M. R\o nnow}
\affiliation{Laboratory for Quantum Magnetism, Institute of Condensed Matter Physics, EPFL, CH-1015 Lausanne, Switzerland}
\author{I. \v Zivkovi\'c}
\email{ivica.zivkovic@epfl.ch}
\affiliation{Institute of Physics, Bijeni\v cka 46, HR-10 000, Zagreb, Croatia}
\affiliation{Laboratory for Quantum Magnetism, Institute of Condensed Matter Physics, EPFL, CH-1015 Lausanne, Switzerland}

\date{\today}
	
\maketitle

%
%
%
%

\textbf{The recent discovery of magnetic skyrmion lattices~\cite{Muhlbauer2009} initiated a surge of interest in the scientific community. Several novel phenomena have been shown to emerge from the interaction of conducting electrons with the skyrmion lattice, such as a topological Hall-effect~\cite{Neubauer2009} and a spin-transfer torque at ultra-low current densities~\cite{Jonietz2010}. In the insulating compound $\coso$, magneto-electric coupling enables control of the skyrmion lattice via electric fields~\cite{White2012,White2014}, promising a dissipation-less route towards novel spintronic devices. One of the outstanding fundamental issues is related to the thermodynamic stability of the skyrmion lattice. To date, the skyrmion lattice in bulk materials has been found only in a narrow temperature region just below the order-disorder transition. If this narrow stability is unavoidable, it would severely limit applications. Here we present the discovery that applying just moderate pressure on $\coso$ substantially increases the absolute size of the skyrmion pocket. This insight demonstrates directly that tuning the electronic structure can lead to a significant enhancement of the skyrmion lattice stability. We interpret the discovery by extending the previously employed Ginzburg-Landau approach and conclude that change in the anisotropy is the main driver for control of the size of the skyrmion pocket. This realization provides an important guide for tuning the properties of future skyrmion hosting materials.}

%
%
%
%

Skyrmions were first introduced in the context of topologically protected excitations in the field theory of hadrons~\cite{Skyrme1962}. Although later the quark theory proved successful, the concept of skyrmions and their mathematical beauty  have found applications in various topics within many body physics, such as 2D electron gases~\cite{Sondhi1993,Brey1995}, Bose condensation~\cite{Ho1998,Khawaja2001} and, particularly, magnetic systems. Initially, magnetic skyrmions were only theoretically predicted~\cite{Bogdanov1994,Rossler2006}, but subsequently they have been found experimentally in the form of skyrmion lattices in several chiral systems, including conductive MnSi~\cite{Muhlbauer2009} and insulating $\coso$~\cite{Seki2012}. On top of the manipulation of the whole skyrmion lattice~\cite{Jonietz2010,White2012,White2014}, successful control of individual skyrmions has been demonstrated using lasers~\cite{Finazzi2013} and scanning-tunneling microscopes~\cite{Romming2013}, indicating 'skyrmionics' to be a promising avenue for the field of spintronics.

\begin{figure*}
	\includegraphics[width=0.9\textwidth]{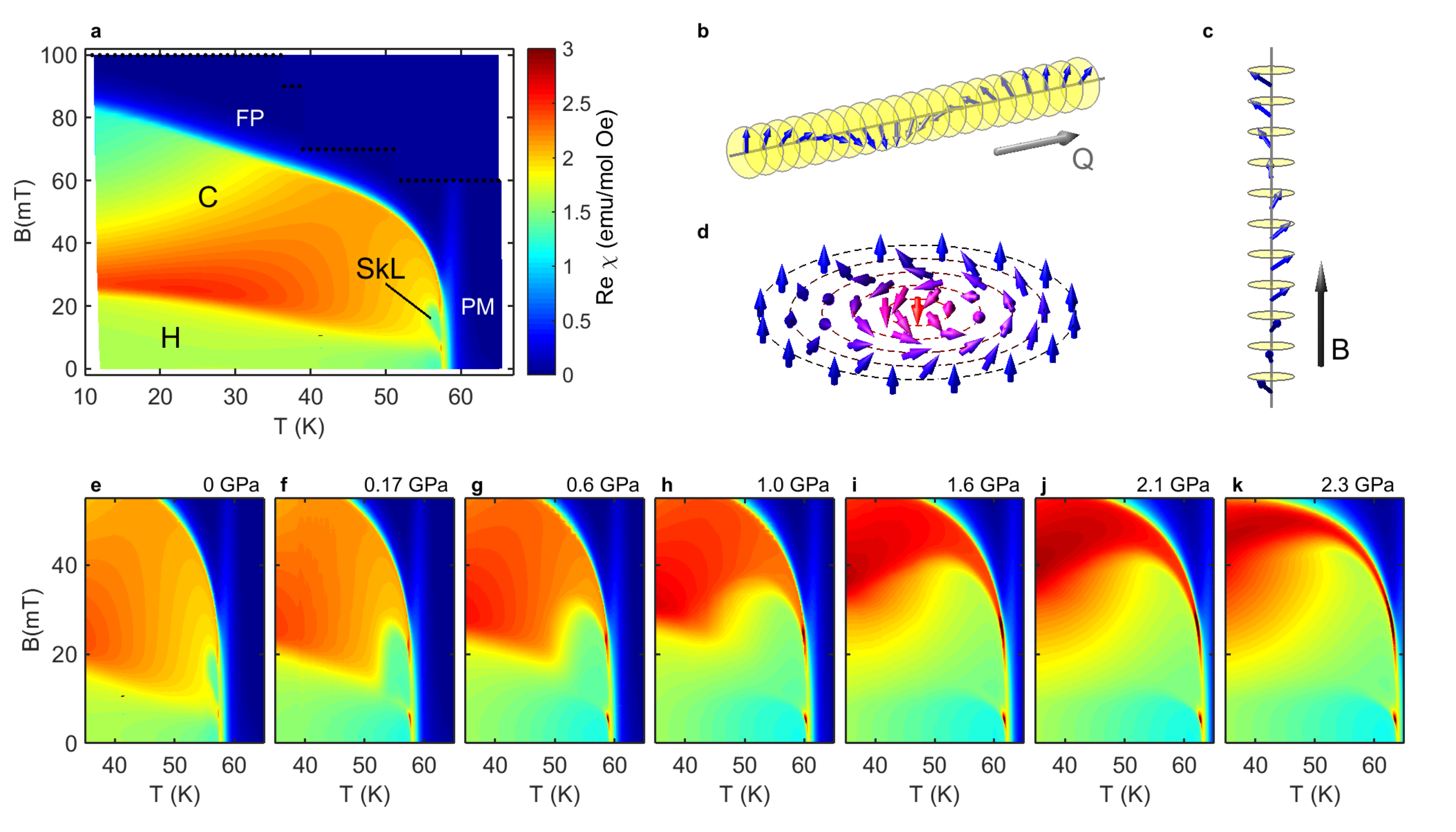}
	\caption{{\bf Development of the phase diagram of $\coso$ with pressure.} {\bf a}: The phase diagram of $\coso$ given by the real part of the magnetic susceptibility $\operatorname{Re} \chi$ taken while ramping magnetic field down. Black dots mark the upper boundary of scans. It consists of H - helical phase, C - conical phase, SkL - skyrmion lattice, FP - field polarized phase, PM - paramagnet. {\bf b}: Helical spin arrangement. Magnetic moments rotate within the plane that is perpendicular to the direction of the wave vector $Q$. {\bf c}: Conical spin arrangement. The precession of magnetic moments outlines the cone which is oriented along the direction of magnetic field $B$. {\bf d}: Spin configuration of a single skyrmion. {\bf e}-{\bf g}: Pressure dependence of the phase diagram around the skyrmion pocket.}
	\label{fig-pocket}
\end{figure*}

The compounds that exhibit a skyrmion lattice display remarkably universal phase diagrams - see for example Figure~\ref{fig-pocket}a. Below a critical temperature $T_C$ the system orders as a helimagnet, with spins rotating within a plane that is perpendicular to the propagation vector $Q$, Figure~\ref{fig-pocket}b. Upon the application of a magnetic field $B > B_{C1}$ the magnetic structure transforms into a conical arrangement, Figure~\ref{fig-pocket}c, where spins precess about a cone that is aligned with the direction of magnetic field. For even larger values $B > B_{C2}$ the spins are field-polarized. The skyrmion phase is found to lie adjacent to the order-disorder transition, at approximately half of the value of the critical magnetic field $B \approx B_{C2}/2$. It exhibits a hexagonal arrangement~\cite{Muhlbauer2009} of individual skyrmions, Figure~\ref{fig-pocket}d, where the central spins point in the opposite direction to the applied magnetic field, and a chiral sense of rotation of the magnetic spins yields a finite topological charge~\cite{Braun2012}.

%
%

So far the temperature width of the skyrmion pocket in bulk samples has been limited to $\Delta T_{SkL}/T_C \sim 3 \%$ in all investigated cubic chiral compounds. A substantial enhancement of the size of the skyrmion pocket has been demonstrated for thin films and ultra thin slabs of these materials~\cite{Yu2010,Seki2012}, with thickness $d \lesssim 100$ nm. In these quasi-2D systems the anisotropy is considered to play a major role in the enhanced stabilization of the skyrmion lattice~\cite{Butenko2010}. On the other hand, the resultant phase diagrams are not generic and depend on the material as well as on the method of preparation~\cite{Li2013,Tonomura2012}. Thus, finding a link between the universal phase diagram in bulk materials and the magneto-crystalline anisotropy presents a major challenge for the deeper understanding of the formation and stability of the skyrmion phase.

In bulk samples the anisotropy $K$ arises in the form of the spin-orbit interaction, and it is the weakest of the energy scales present in skyrmion compounds $K \ll D \ll J$, where $D$ and $J$ are the Dzyaloshinskii-Moriya and the ferromagnetic exchange interactions, respectively. 
$D$ favors a perpendicular orientation of neighboring spins, and is responsible for the formation of the chiral helimagnet ground state with a long wavelength modulation $\lambda \sim J/D \gg a$, $a$ being the inter-atomic distance. The anisotropy $K$ then pins the helices along preferred directions, and hence determines the value of the first critical magnetic field $B_{C1}$ above which the conical structure is stabilized.

To further our understanding of the stability of the skyrmion lattice, it is imperative to establish the parameter range in which the skyrmion lattice is physically favorable with respect to other phases, and further how the size of the skyrmion phase depends on $J$, $D$ and $K$. In that context, the application of hydrostatic pressure is a well-known technique which allows fine-tuning of energy scales through tiny shifts in atomic positions. Before the discovery of its skyrmion lattice phase, MnSi was intensively investigated~\cite{Pfleiderer2001,Doiron2003} due to a complete suppression of long-range magnetic order above a critical pressure ($p_C \sim 1.5$ GPa) and the discovery of non-Fermi liquid behavior for $p > p_C$. A similar suppression of magnetic order has been observed in FeGe~\cite{Pedrazini2007} around 19 GPa. In contrast, the ordering temperature of $\coso$ increases under pressure~\cite{Huang2011,Sidorov2014}, thus emphasizing the importance of detailed investigation of its pressure dependence. Here we present an extensive study of the phase diagram of $\coso$ under hydrostatic pressure up to 2.3 GPa, and discover that the absolute size of the skyrmion pocket in bulk samples can be dramatically enlarged.

High-quality magnetic ac susceptibility $\chi = \operatorname{Re} \chi + i\operatorname{Im} \chi$ accurately traces the magnetic phases, as illustrated in Figure~\ref{fig-pocket}a for zero-pressure. As can be seen from the zoomed-in part in Figure~\ref{fig-pocket}e, the skyrmion phase is manifested as a region of lower susceptibility (yellow) compared to the surrounding conical phase (red). For $p = 0$ it occupies a very small part of the phase diagram, adjacent to the order-disorder boundary, with a maximum extent in temperature of 2 K. The maximum field range amounts to around 15 mT, and at lower field values the skyrmion pocket becomes narrower, forming a shape of an inverted tear-drop. The situation drastically changes by the application of even a small hydrostatic pressure ($p = 0.17$ GPa, Figure~\ref{fig-pocket}f). The skyrmion pocket increases in size, especially the low magnetic field region, more than doubling over the zero-pressure extent. In addition to the observed growth, one can notice a qualitative change in the way the phases are arranged in that part of the phase diagram. Namely, under pressure the skyrmion phase borders directly with the helimagnetic phase without the conical phase in-between. As shown in Figures~\ref{fig-pocket}g-k, further increase of the pressure evidences a continuation of the growth of the low-susceptibility region, both towards lower temperatures as well as towards larger magnetic fields.

%
\begin{figure*}
	\includegraphics[width=0.9\textwidth]{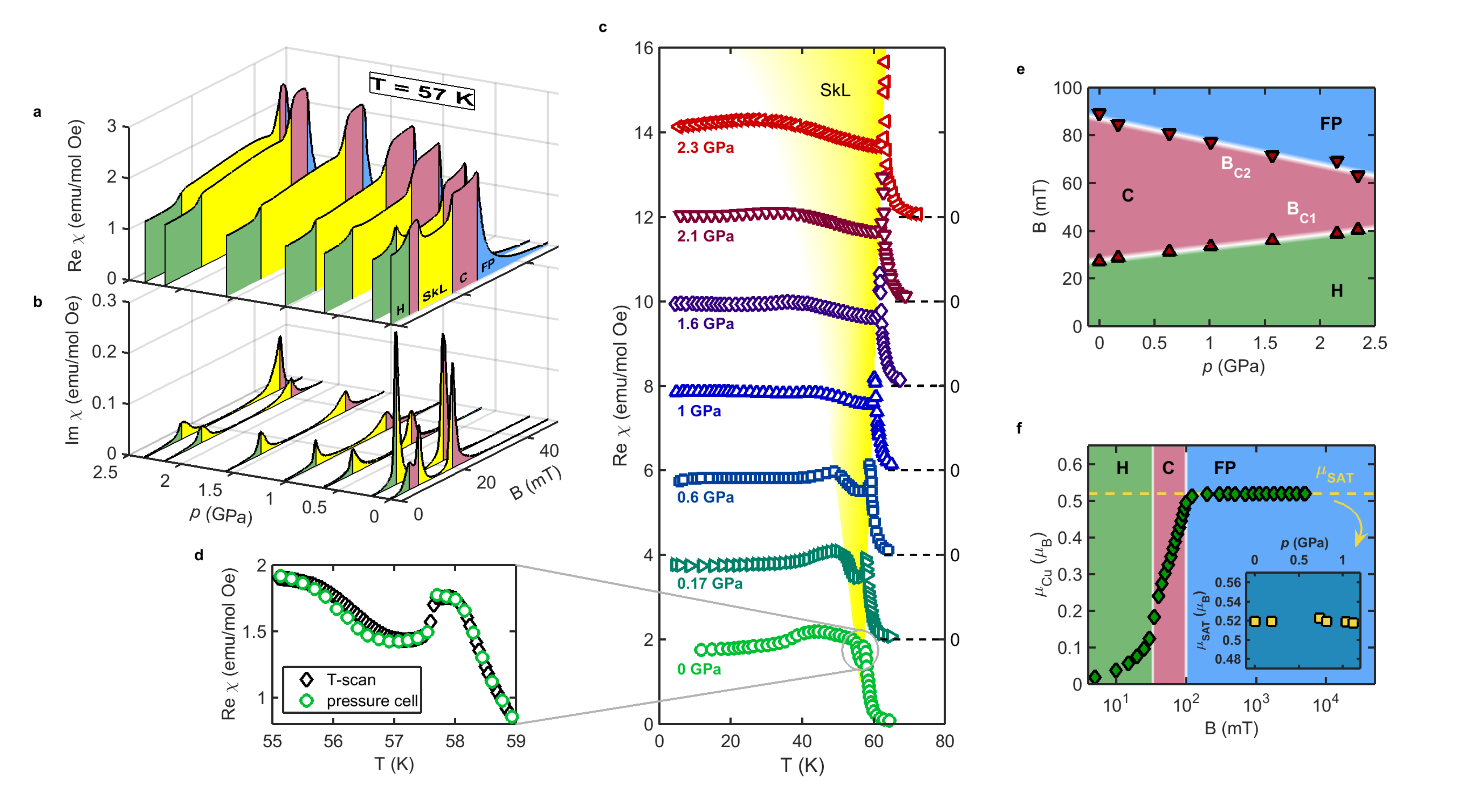}
	\caption{{\bf Evolution of magnetic phases with pressure.} {\bf a}: real part and {\bf b}: imaginary part of the magnetic susceptibility $\chi$ at $T = 57$ K. {\bf c}: Extracted temperature profiles through the middle of the skyrmion pocket at magnetic fields $B = 15, 16.5, 18, 19.5, 21, 22.5$ and 24 mT (from 0 to 2.3 GPa). {\bf d}: Comparison of the extracted temperature profile with a temperature scan taken in cooling measured on the same sample using a standard setup (without the pressure cell) at zero pressure and $B = 15$ mT. A small difference is visible on the low temperature side of the skyrmion pocket ($\leq 0.2$ K) where disintegration of the skyrmion lattice occurs and the exact path through the phase diagram becomes important. {\bf e}: Pressure dependence of the helical-conical and conical-field polarized metamagnetic phase transitions in the $T = 0$ K limit. {\bf f}: Magnetic field dependence of magnetization at $p = 1.0$ GPa and $T = 10$ K. Inset: Pressure dependence of the saturation magnetization level.}
	\label{fig-phases}
\end{figure*}
%

Quantitatively, we can identify the phase boundaries and extract their pressure dependence from individual scans. Figures~\ref{fig-phases}a-b show field scans of real and imaginary susceptibility at $T = 57$ K, the middle of the skyrmion pocket at zero pressure. The phase boundaries are determined by maxima in the imaginary component, except for the transition between the conical (red) and the field polarized state (blue) where a kink in the real part marks the boundary. On the high-field side the skyrmion pocket (yellow) transforms into the conical phase by a steep increase of the real component, while the imaginary part exhibits a sharp peak. On the low-field side the presence of the conical phase between the helimagnetic (green) and the skyrmion phase at zero pressure is revealed by two peaks in the imaginary part, while at elevated pressures a single peak is observed. To our knowledge this represents the first experimental evidence that the helimagnetic phase and the skyrmion phase are thermodynamically distinct. In the real part this transition becomes less pronounced, because $T_C$ increases with pressure, placing field scans at 57 K further away from the ordering temperature. We showed recently that the strength of the susceptibility anomaly marking the border of the skyrmion pocket becomes quickly suppressed for $T < T_C$~\cite{Levatic2014}.

In our experiment the temperature was slowly increased while magnetic field was continuously ramped between zero and a maximum value. Due to the slow temperature change ($dT/dt < 100 \mu$K/s) and reduced field range around the skyrmion pocket (black dots in Figure~\ref{fig-pocket}a), successive field scans were spaced by no more than 0.2 K, providing a very good resolution in temperature. This enables us to determine the temperature width of the skyrmion phase by plotting $\chi (T,B)$ for fixed $B$. Extracted temperature profiles through the middle of the skyrmion pocket are displayed in Figure~\ref{fig-phases}c. In Figure~\ref{fig-phases}d we compare $\chi (T)$ extracted from the $\chi (T,B)$ maps to a temperature scan recorded on cooling using a standard susceptibility setup. The overlap of the two curves demonstrates that the extracted profiles presented in Figure~\ref{fig-phases}c probe directly the thermodynamic phase boundaries.

Even at the smallest applied pressures an enhanced stability of the skyrmion phase is readily observed: compared to the zero pressure case, where $\Delta T_{SkL} \approx 2$ K (3\% of $T_C$), it reaches $\Delta T_{SkL} \approx 10$ K at 0.6 GPa, corresponding to 17\% of $T_C$. If the same relative increase is achieved in the recently discovered skyrmion compound~\cite{Tokunaga2015} Co$_8$Zn$_9$Mn$_3$ with the ordering temperature at 322 K ($49^{\circ}$C), the skyrmion lattice would be stable even below $0^{\circ}$C, this covering the usual operational temperature range for most electronic circuits.

With further increase of pressure the low-temperature boundary to the conical phase becomes less pronounced, transforming into a wide cross-over. At the highest pressure ($p = 2.3$ GPa) the low susceptibility region extends almost to 30 K, indicating that skyrmions exist down to at least half of the ordered phase diagram in the high-magnetic field region. It remains an open question as to what kind of a spatial distribution is taken by skyrmions at high pressures and whether they still form a well-defined lattice.

At low temperatures the helical and conical phases are thermodynamically stable, although with pressure the area covered by the conical phase is substantially reduced from both high and low field sides compared with ambient pressure, as revealed in Figure~\ref{fig-phases}e. More importantly, we observe an increase of the first critical magnetic field $B_{C1}$ that marks the transition from the helical to the conical phase. As has been introduced above, $B_{C1}$ is linked directly to the anisotropy energy $K$ which determines the direction of the helical propagation vector in the absence of an applied magnetic field $B$. Through the adaptation of the effective model developed in Ref~\onlinecite{Muhlbauer2009}, below we demonstrate that such an increase of $K$ can explain the observed expansion of the bulk skyrmion phase stability.

We start from the usual Ginzburg-Landau functional for the magnetization $\textbf{M}$ of the system~\cite{Izyumov1984}

\begin{equation}
\label{eq-functional}
F[ {\mathbf M} ]= A  {\mathbf M}^2 
+ J({\nabla \mathbf M})^2 + D{\mathbf M} \cdot (\nabla \times {\mathbf M}) + K{\mathbf M}^4 - {\mathbf B} \cdot {\mathbf M},
\end{equation}

where the ${M}^2$ term is related to the temperature-dependent tuning parameter $\tau (T) = 1 - A(T)J/D^2$. Here, $A(T)$ controls the transition and, in our approach, is adopted to reproduce the experimental $T_C(p)$ and enable us to predict the phase diagram in absolute units of temperature. The microscopic parameters $J$, $D$ and $K$ are related to experimental observables in a simple way: $T_C \propto J$, $B_{C2} \propto D^2/J$ (Ref.~\onlinecite{Janson2014}), while as mentioned previously we assume $B_{C1} \propto K$.

\begin{figure}
	\includegraphics[width=0.45\textwidth]{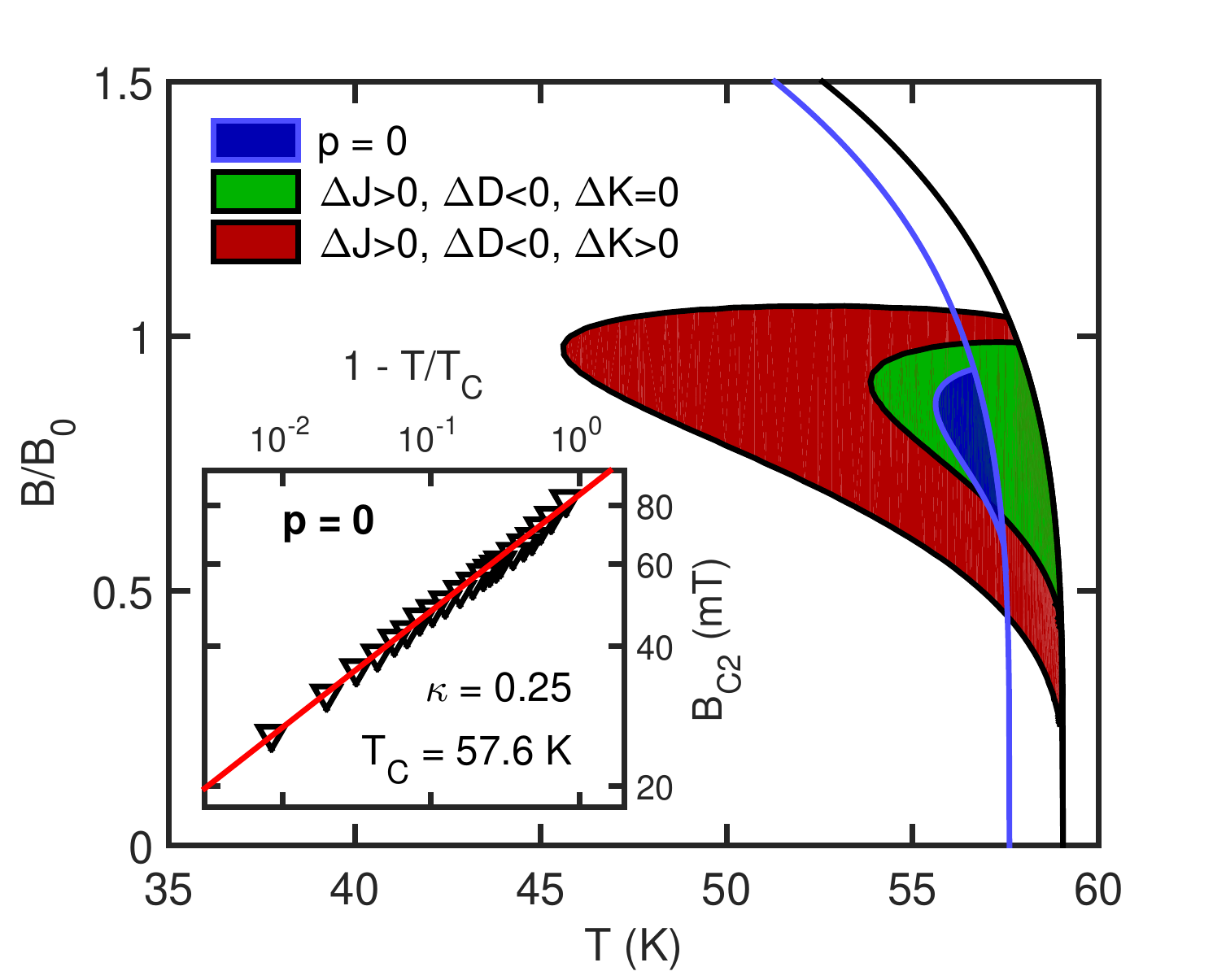}
	\caption{{\bf Modeling of the size of the skyrmion pocket using the Ginzburg-Landau approach}. The width of the calculated skyrmion pocket ($\sim 2$ K) for $p = 0$ has been used to fix the values of all parameters within the model, see Methods for details. To simulate the behavior at elevated pressures, we change the values of $J$, $D$ and $K$ as obtained at $p = 0.6$ GPa: $\Delta J / J = 2.5$ \%, $\Delta K / K = 15$ \% and $\Delta D / D = -3.5$ \%. The size of the skyrmion pocket is significantly increased only if the change of anisotropy is incorporated in the calculation. Inset: critical behavior of the upper metamagnetic transition between the conical and field-polarized phase $B_{C2} \sim (1 - T/T_C)^{\kappa}$.}
	\label{fig-model}
\end{figure}

It has been shown~\cite{Muhlbauer2009} that when thermal fluctuations around the mean-field solution are included, the skyrmion lattice phase becomes stable close to $T_C$. However, it remained unknown as to the width of the temperature window over which the stabilization can be expected to occur. In order to address this question, first it is necessary to establish the functional relation between the tuning parameter $\tau$ and the thermodynamic temperature $T$. Around the mean-field solution the upper-critical field $B_{C2}$ is related to $\tau$ through $B_{C2}(\tau) \cong \sqrt{2(\tau - \tau_0)}$, where $\tau_0$ reflects the shift of the transition temperature away from its mean-field value due to the linear effects of fluctuations. In addition, $B_{C2}(T)$ can itself be directly extracted from the experimentally established phase diagram, Figure~\ref{fig-pocket}a. We find that it follows the form of a critical behavior

\begin{equation}
\label{eq-kappa}
B_{C2}(T) = z_{T}(1 - T/T_C)^{\kappa},
\end{equation}

with critical exponent $\kappa \cong 0.25$, as presented in the inset of Figure~\ref{fig-model}. This leads to the simple relation for $\tau (T)$

\begin{equation}
\label{tau}
\tau (T) - \tau_0 = z_T (1 - T/T_C)^{2\kappa} = z_T \sqrt{1 - T/T_C}.
\end{equation}

The square-root dependence is preserved across the pressure range investigated in this study.

%
%

Now we can proceed with the discussion of the size of the skyrmion pocket in terms of thermodynamic temperature $T$. As mentioned above, the inclusion of thermal fluctuations is necessary to stabilize the skyrmion lattice over the conical spin arrangement. However, it has been suggested recently~\cite{Janoschek2013} that the fluctuations close to $T_C$ are strongly interacting, lowering the transition temperature and making it first-order. Within the effective model the reduction in $T_C$ is implemented~\cite{Muhlbauer2009} by setting the corrections of the order parameter to be smaller than 20\%. We keep this approach and in the main panel of Figure~\ref{fig-model} plot the resultant order-disorder boundary with the outline of the skyrmion pocket for different values of $J$, $D$ and $K$. For the zero-pressure case we adjust the coefficients of the model so that the skyrmion pocket is limited to $\sim 2$ K below $T_C$, as found in the experiment. To explore the effect of pressure we consider the particular case of $p = 0.6$ GPa, where the skyrmion pocket is still well defined, and estimate $\Delta J$, $\Delta D$ and $\Delta K$ from the experimental data: $\Delta T_C / T_C = \Delta J / J = 2.5$ \%, $\Delta B_{C1} / B_{C1} = \Delta K / K = 15$ \%, while $\Delta B_{C2} / B_{C2} = -9$ \% translates to $\Delta D / D \approx -3.5$ \%. It can be shown that the skyrmion pocket grows if the parameter $JK/aD^2$ increases, indicating that all of the observed changes contribute positively. If we take into account only $\Delta J$ and $\Delta D$ ($\Delta K = 0$) then we obtain a small increase of the skyrmion pocket. However, a more significant growth is obtained when the observed $\Delta K$ is included, giving $\Delta T_{SkL} \sim 10$ K. The temperature extent is here only indicative since it is known that the Ginzburg-Landau approach is not valid far away from $T_C$. Additionally, $B_{C1} \propto K$ is only a crude approximation which is expected to hold only in a narrow range of parameters. Nevertheless, our analysis demonstrates the importance of anisotropy in stabilizing the skyrmion lattice at temperatures far from the immediate proximity of the ordering temperature. Indeed just $\sim 15\%$ change in effective anisotropy can expand the temperature range of the skyrmion stability by 600\%!

The crucial role of anisotropy has also been explored in Monte Carlo simulations~\cite{Buhrandt2013} where it has been found that the correct phase diagram can be reproduced if additional anisotropy-compensation terms are added because of boundary effects. On the experimental side, in the recently discovered skyrmion compound GaV$_4$S$_8$ the orientation of skyrmions is dictated by the magnetic easy axis and not by the direction of magnetic field~\cite{Kezsmarki2015}. At the same time, the relative size of the skyrmion pocket within the phase diagram of this polar magnet is substantially larger than in the family of chiral magnets such as MnSi and $\coso$.

While the above-presented model is applied as a continuous-field approximation, an important question is what qualitative changes occur with pressure on the level of the unit cell. This is especially important for $\coso$ since there are two crystallographically distinct copper sites which couple strongly into tetrahedra with a quantum triplet ground state~\cite{Janson2014}. These triplets form a trillium lattice which is identical to the lattice of Mn ions in MnSi, and long-range order is then governed by the effective exchange and Dzyaloshinskii-Moriya interactions between the triplets. It follows that the effect of pressure on $\coso$ can be two-fold: (i) changing the quantum nature of tetrahedrons and (ii) changing the inter-triplet interactions. In order to probe the former effect, we have performed measurements of the magnetization plateau under pressure. The plateau occurs when the system enters the field polarized state for magnetic fields $B > 100$ mT, see Figure~\ref{fig-phases}f and inset. It turns out that within the error-bar the observed magnetization plateau does not change, at least for pressures $p \lesssim 1$ GPa. This means that the tetrahedra remain relatively rigid, while pressure-induced shifts in atomic positions occur dominantly between the tetrahedrons, influencing the effective inter-triplet interactions and justifying the use of the continuous model.

We believe that the results presented in this study and the supporting model clearly demonstrate that although it is the weakest of the energy scales, anisotropy plays a decisive role in stabilization of the skyrmion lattice. Broadening the operational range for the skyrmion lattice and making it more resistant against external fluctuations provides a promising route for future investigations in order to enhance the potential of other skrymion-supporting materials, especially around room temperature~\cite{Tokunaga2015}, where skyrmion-based applications are foreseen. The deeper theoretical understanding of the various mechanisms leading to skyrmion stability may also guide the rational design of new compounds with favorable characteristics for hosting stable skyrmion phases.

%
%
%
%

\section{Methods}
\label{Methods}

{\bf{Experimental conditions}}.
The single crystal sample was prepared by the chemical vapor transport technique. It was aligned by x-ray Laue backscattering and cut along the $\left[111\right]$ direction with dimensions $4 \times 1 \times 1\:$mm$^3$. The ac susceptibility measurements were performed using a balanced coils setup. Two detection coils were wound on an inox tube 2 mm in diameter, with centers of the coils 2.5 mm apart. On top of each detection coil a driving coil was made for generation of ac magnetic field. AC current was supplied by a Keithley 6221 current source and the balance of the coils was achieved by changing a resistance in series with each driving coil. The amplitude of ac magnetic field was 0.1 mT. The detection of the ac signal was done by the lock-in amplifier Signal Recovery 7265. The measurements of the phase diagram for each pressure were performed in a slow temperature drift regime, where the sweeping rate has been controlled between 100 $\mu$K/s around the skyrmion phase and 700 $\mu$K/s below $\sim$ 30 K. The magnetic field was generated by a Cryomagnetics 9 T magnet with a sweep rate of 0.1 mT/s (the slowest available on the instrument) between zero and a maximum field value of $B = 100$ mT. At higher temperatures the upper field value was regularly decreased in order to increase the density of the measured points inside the ordered part of the phase diagram.

The pressure for AC susceptibility study was generated using a non-magnetic piston cylinder pressure-cell with Daphne oil 7373 as a transmitting medium. Pressure was determined using the relative change of the resistance of a manganine wire compared to the zero-pressure case. The magnetization under pressure was measured in a Quantum Design SQUID magnetometer MPMS using a commercial EasyLab pressure cell. The sample used was $1 \times 1 \times 1\:$mm$^3$ taken from the same batch.

{\bf{Ginzburg-Landau phenomenology}}.
The phase diagram is expected to be qualitatively captured by Ginzburg-Landau theory for modulated magnetic structures. For the detailed description of the approach we refer to the Refs. \onlinecite{Muhlbauer2009,Izyumov1984} and just highlight here the main features of the formalism. The calculations are made in the continuous-field approximation justified by the large ratio of the skyrmion radius $\lambda$ to the lattice constant $a$. The Ginzburg-Landau functional is taken up to fourth order in magnetization and second order in magnetization gradients, thus including both the Dzyaloshinskii-Moriya interaction and anisotropies. The magnetic structure is determined by minimization of the free energy, where the order parameter is naturally taken as the local magnetization reduced by the average magnetization of the crystal. A mean-field treatment shows the conical phase to be energetically favorable, but the skyrmion phase lies only slightly higher in energy. By including the Gaussian fluctuations of the free energy the skyrmion phase becomes favorable in a certain range of finite magnetic fields. The fluctuations contribute mainly at the short-length scale and are calculated with the cut-off in the momentum space in order of $2 \pi/a$. The transition to the paramagnetic state is expected to occur when the fluctuations become significant (around $20 \%$ of the mean-field value), which determines the ordered phase boundary through the relation $B_{C2}(\tau) = \sqrt{2(\tau - \tau_0)}$. For the calculation of the phase diagram at zero pressure we used the following parameters: $\gamma = JD/K = 6.2$, $\Lambda = \lambda/a = 28$, $z_T = 3.4$ and $\tau_0 = 3.1$. The values of $\gamma$ and $\Lambda$ are in good agreement with a recent density functional calculation~\cite{Janson2014}. The Ginzburg-Landau approach is considered to break down away from $T_C$ so we leave the field axis normalized to $B_0 = \sqrt{(JQ^2)^3/K}$.

%
%
%
%
%

\section{Acknowledgement}
\label{Acknowledgement}

We acknowledge discussions with A. Tsirlin, O. Janson, I. Rousochatzakis, N. Nagaosa and A. Rosch. The support from Croatian Science Foundation Project No. 02.05/33, Croatian Ministry of Science, Education and Sport No. 035-0352826-2848, Swiss National Science Foundation and ERC project CONQUEST and Austrian Science Foundation project P27980 - N36 are acknowledged.

\section{Author contributions}
\label{Contributions}

I.\v Z. planned the project. I.L., P.P., V.S. and I.\v Z. performed the measurements. I.L., V.S. and I.\v Z. analyzed the data. H.B. and A.M. performed the sample preparation. The theoretical modeling was carried out by A.K. A.K., J.S.W., H.M.R. and I. \v Z. wrote the manuscript.


\end{document}